# Fluctuation, insulation and superconductivity: the pressure-dependent phase-diagram of $Rb_2Mo_6Se_6$

Yongsheng Zhao[1,2], Alexander P. Petrović[3,†], Patrick Gougeon[4], Philippe Gall[4], Chen Kai[5], Wenge Yang[1,*] and Moritz Hoesch[2,*]

[1] Center for High Pressure Science and Technology Advanced Research (HPSTAR), 1690 Cailun Road, Shanghai 201203, P. R. China
[2] Deutsches Elektronen-Synchrotron DESY, Notkestrasse 85, 22607 Hamburg, Germany
[3] Department of Quantum Matter Physics, University of Geneva, 24 Quai Ernest-Ansermet, 1211 Geneva, Switzerland
[†] Present Address: Division of Physics and Applied Physics, School of Physical and Mathematical Sciences, Nanyang Technological University, Singapore 637371.
[4] Institut des Sciences Chimiques de Rennes, UMR 6226 CNRS – Université de Rennes 1 – INSA de Rennes, 11 Allée de Beaulieu, CS 50837, 35708 Rennes Cedex, France
[5] National Synchrotron Radiation Laboratory, University of Science and Technology of China, Hefei 230026, Anhui, China

[*] Corresponding Authors:
yangwg@hpstar.ac.cn
moritz.hoesch@desy.de

# Abstract

The quasi-one-dimensional ($q1D$) material $Rb_2Mo_6Se_6$ has been proposed to display a nontrivial combination of low-dimensional fluctuations and a dynamical charge density wave (CDW) at ambient pressure. This may lead to a progressive metal to insulator cross over at low temperature. To explore the link between the crystal dimensionality and this insulating instability, we have performed hydrostatic pressure-dependent electrical transport measurements on single crystals of $Rb_2Mo_6Se_6$. At low pressure, we observe thermally-activated behavior consistent with a temperature-dependent gap $E_g(T)$ opening below a characteristic temperature $T_{R\min}$. Upon increasing the pressure $T_{R\min}$ initially rises, indicating a reinforcement of the low temperature insulating state despite a continuous reduction in $E_g(P)$. We interpret this as a signature of suppressed fluctuations as the dimensionality of the electronic structure rises. However, $T_{R\min}$ drops above 8.8 GPa and superconductivity emerges at 12 GPa. Between 12-24.2 GPa the superconducting and insulating instabilities coexist, with superconductivity surviving up to the maximum attained pressure (52.8 GPa). Analysis of the magneto transport reveals two distinct regions: at high pressures the anisotropy gradually falls and the superconducting state appears unremarkable. In contrast, coexistence with the gapped insulating phase creates a superconducting dome. The emergence of a peak in the critical temperature $T_c$ despite the depleted density of states is indicative of enhanced coupling. Our journey from the extreme 1D to 3D limits in this prototypical q1D metal reveals an intriguing relationship between superconducting and insulating ground states which is simultaneously competitive and symbiotic.

# Introduction

The electronic structure of quasi-one-dimensional (q1D) materials is characterized by almost complete confinement of the metallic charge carriers to propagation along the chains. Few materials accommodate such extreme q1D behaviour [1]. The Fermi surface of these metals consists of nearly flat sheets from bands that only disperse along the chain direction, in the limiting case of completely flat sheets characterized by a single Fermi wave number $k_F$. These sheets will easily facilitate Fermi surface nesting, *i.e.* a common spanning vector $q = 2k_F$, thus leading to an instability of the material towards the formation of a charge density wave (CDW) ground state through the Peierls mechanism [2]. The corresponding phase transition manifests itself in the form of a resistivity anomaly, a periodic lattice distortion which may be observed by diffraction and/or a spectroscopically observable gap in the electronic structure. A dramatic consequence of one-dimensional confinement of the free charges is the instability of the Fermi liquid and its quasiparticles towards excitations that are necessarily collective in their nature. In real q1D materials, this so-called Tomonaga Luttinger Liquid (TLL)[3] state is restricted to a rather high temperature regime in which the electronic wavefunctions cannot propagate coherently between the chains [4, 5]. At low temperature (*T*), the highly nested Fermi surfaces of q1D materials lend themselves to the formation of lattice distortions due to the Peierls mechanism [1]. Peierls transitions can obfuscate observation of the characteristic scaling of a TLL and its eventual dimensional cross-over. Clarifying the existence of a Peierls ground state, determining its transition temperature, and detecting the associated gap in the electronic structure are hence vital to the interpretation of experimental data in q1D metals [6]. In this paper we set out to clarify the existence of a Peierls effect in a particularly anisotropic one-dimensional material whose complex low-*T* characteristics are indicative of an emergent insulating instability.

The family of cluster chain compounds $M_2Mo_6X_6$ (*M* = group IA or group III metal, *X* = S, Se or Te) features infinite-length crystallographic chains $(Mo_3X_3)_\infty$, doped by electrons from the *M* ions [7, 8]. Charge transport is highly anisotropic, with electron flow parallel to the chains greatly exceeding the transverse hopping ($t_\parallel \gg t_\perp$) [9]. The conduction electrons occupy a single broad, linearly-dispersing band of predominant Mo *d* character, which generates an ideal q1D Fermi surface consisting of two minimally-warped sheets at $\pm k_F$ [10]. This family of materials [7, 8] is hence an ideal laboratory to explore the impact of varying dimensionality on the *T*-dependent charge dynamics as well as the electronic ground state [11-19], which has long been suspected to be CDW-like for *M* = K, Rb, Cs [18, 20, 21]. No periodic lattice distortion has thus far been observed by X-ray measurements, nor has there been any observation of a structural transition in any family member. Spectroscopic signatures of a CDW gap are also absent from the literature. Some members of the $M_2Mo_6X_6$ family have shown experimental signatures that are compatible with a low-*T* state linked to a TLL [10, 22, 23]. The *M* = Tl compound has also shown dielectric instabilities at low *T* [24]. In strained $Tl_2Mo_6X_6$ crystals, the observation of broadband transport noise [13] gives the strongest hint at CDW physics so far, along with an intriguing transition from a superconducting instability to a "bad metallic" ground state displaying a low-$T$ resistivity upturn when strained [12].

Here, we focus on Rb$_2$Mo$_6$Se$_6$ in which the combination of a large inter-chain spacing and weak hopping via the Rb valence $s$ orbitals results in a particularly low $t_\perp$ ~ 30 K [20]. The electrical resistivity of Rb$_2$Mo$_6$Se$_6$ is characterized by metallic behavior at room $T$, a resistivity minimum around $T_{R\min}$ ~170 K and a strong upturn in the resistivity at low $T$ [11, 20, 25]. Similar insulating characteristics at low $T$ in the q1D metals Ta$_2$Se$_8$I [26], Mo$_4$O$_{11}$ (under pressure) [27], the purple bronze Li$_{0.9}$Mo$_6$O$_{17}$ [28, 29] and Ta$_{1.2}$Os$_{0.8}$Te$_4$ [30] have been assigned to emergence of a CDW, while upturns in PrBa$_2$Cu$_4$O$_8$ [31] and Na$_{2-\delta}$Mo$_6$Se$_6$ [32] are instead consistent with strong localization.

In this paper, we report high pressure electronic transport measurements on Rb$_2$Mo$_6$Se$_6$ single crystals to study the evolution of the metal-insulator transition. At low pressure, we find that although Rb$_2$Mo$_6$Se$_6$ is metallic at high $T$, a resistivity upturn below $T_{R\min}$ is well-described by the opening of a $T$-dependent gap $E_g(T)$, which unusually exhibits a finite-$T$ maximum before falling at the lowest $T$. The isothermal $E_g(P)$ decreases monotonically with rising pressure, whereas an initial rise in $T_{R\min}$ occurs prior to the suppression of the insulating phase. A superconducting (SC) ground state emerges at higher pressures, in agreement with early experiments on polycrystalline samples [25, 33]. Extending our measurements to pressures more than 3 times higher than previously attained, we trace out a SC dome which coexists with a partially gapped Fermi surface, before smoothly evolving into an increasingly isotropic conventional superconductor in the high-pressure limit.

## Experiments

### Syntheses and Crystal Growth

The starting materials used for the solid-state synthesis and crystal growths were InSe, MoSe$_2$, RbCl, and Mo, all in powder form. These were kept and handled in a purified argon-filled glovebox. Before use, the Mo powder was reduced under H$_2$ flowing gas at 1000°C for ten hours in order to eliminate any trace of oxygen. Powder samples of the Rb$_2$Mo$_6$Se$_6$ compound were prepared in two stages. The first step was the synthesis of In$_2$Mo$_6$Se$_6$ from a stoichiometric mixture of InSe, MoSe$_2$ and molybdenum heated at 1000°C in an evacuated sealed silica tube for 36 hrs. The second step was an ion exchange reaction of In$_2$Mo$_6$Se$_6$ with RbCl at 800°C. For the latter process, powders of In$_2$Mo$_6$Se$_6$ and RbCl in a ratio of approximately 1:2.5 were mixed and then cold pressed. The pellet was subsequently sealed under vacuum in a long silica tube. The end of the tube containing the pellet was placed in a furnace while the other end was about 5 cm out from the furnace, at room temperature. The part of the tube containing the pellet was heated at 800°C for 48 h. After reaction, white crystals of InCl were observed at the cool end of the tube. The resulting product was found to be single-phase on the basis of its powder X-ray diffraction diagram made on a D8 Bruker Advance diffractometer equipped with a LynxEye detector (Cu Kα₁ radiation). Single crystals of Rb$_2$Mo$_6$Se$_6$ were obtained by heating cold pressed powder samples (~ 5g) in a molybdenum crucible sealed under a low argon pressure using an arc-welding system. The charge was heated at 300 °C/h up to 1500 °C, then held for 3 hours, before cooling at 100 °C/h down to 1000°C and finally furnace cooled. Crystals thus obtained have the shape of needles with hexagonal cross-sections [7].

## Electronic Transport Measurements

We measured the electrical resistance using a standard DC four-probe method in an exchange gas cryostat with thermal stability ± 1mK below 10 K. A Rb$_2$Mo$_6$Se$_6$ crystal was contacted with four probe wires (Pt, 4 μm diameter, using conductive Epoxy silver paste CW2400 from ITW Chemtronics), then loaded into a cubic boron nitride (CBN) insulator chamber in a diamond anvil cell (DAC). An additional four Au probes were pasted to the surface of the CBN for electrical connections to the sample. The pressure was calibrated by the ruby luminescence method [34]. Silicon oil was used as the pressure-transmitting medium, with the highest achieved pressure being 52.8 GPa. The excitation current was applied parallel to the crystallographic $c$ axis, which was oriented perpendicular to the external magnetic field.

## Results and Analysis

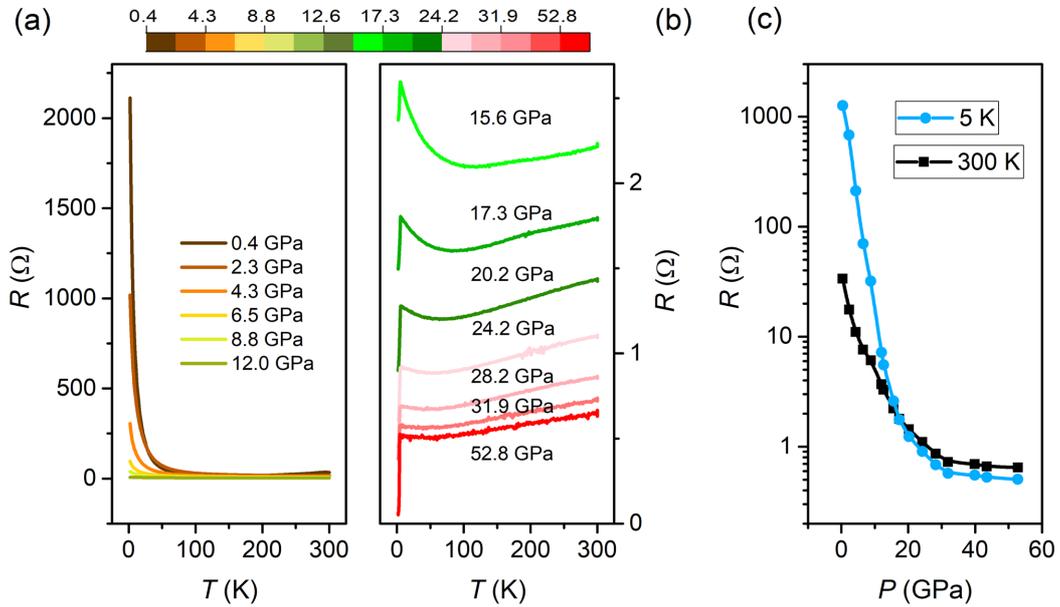

FIG. 1. Temperature dependence of the electrical resistance in single crystal Rb$_2$Mo$_6$Se$_6$ at various pressures: (a) 0.4 GPa-12.0 GPa; (b) 15.6 GPa-52.8 GPa, where the superconducting phase is observed. All isobaric data in this manuscript have been plotted using the same color scale to indicate the pressure at which they were acquired. (c) Pressure evolution of the isothermal electrical resistance at $T$ = 300 K and 5 K.

Isobaric measurements of the $c$ axis electrical resistance $R(T)$ are shown in Fig.1 (a,b),

together with the extracted isothermal $R(P)$ for $T = 5$ K and 300 K [Fig.1 (c)]. Upon cooling at low pressure, $R(T)$ exhibits metallic behavior before passing through a broad minimum at temperature $T_{Rmin}$. Although $R(P)$ falls with increasing pressure at all temperatures, the upturn at low $T$ is preferentially suppressed [Fig.1(c)]. Indeed, $R(P)$ drops by over three orders of magnitude at $T = 5$ K, whereas $R(P)$ decreases by less than 2 orders of magnitude at $T = 300$ K over the same pressure range. This already indicates that small pressures enhance the intrachain hopping of the compounds and hence destabilize the insulating state. Above 12 GPa a SC phase emerges [Fig.1(b)], which coexists with a clear insulating tendency in $R(T)$ till 28.2 GPa. Above this value, the upturn in $R(T)$ is extinguished, but the SC phase survives up to at least 52.8 GPa (the maximum pressure achievable in our apparatus).

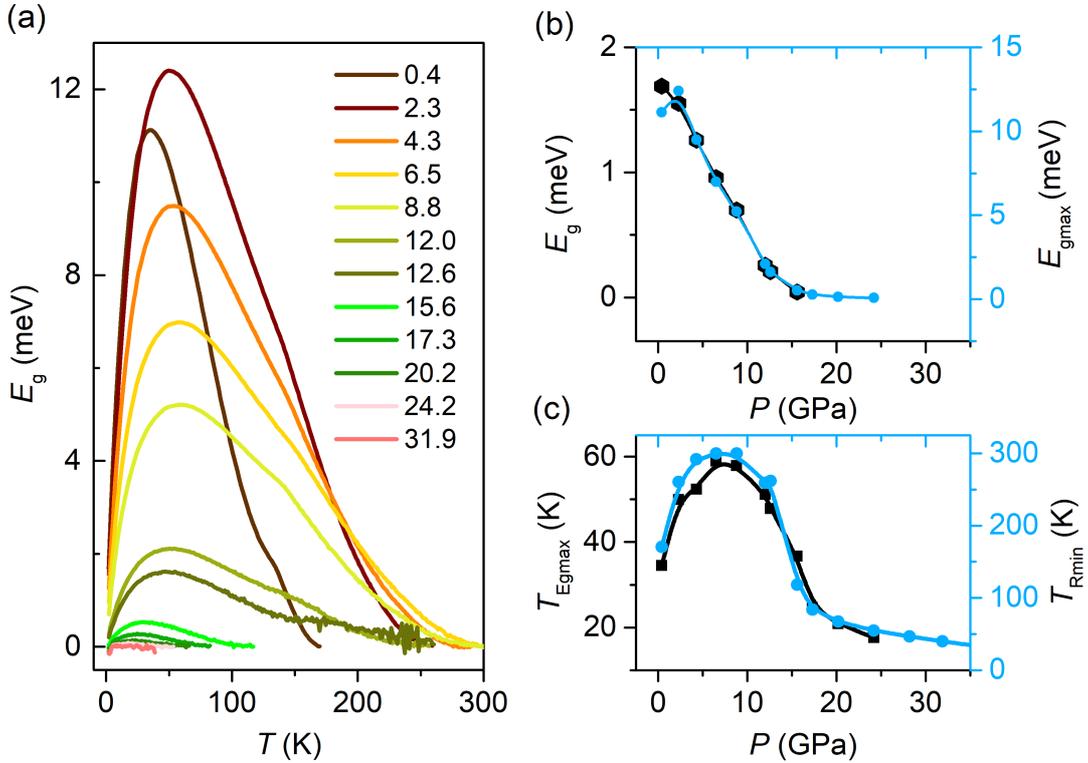

FIG. 2. (a) Temperature dependence of the gap magnitude $E_g(T)$ extracted according to equation (1) in the text. (b) Evolution of $E_g$ ($T = 2$ K) and $E_{gmax}$ with pressure. (c) Variation of $T_{Egmax}$ and $T_{Rmin}$ with pressure, revealing a broad maximum between 5-10 GPa.

Several possible mechanisms in q1D materials could induce an upturn in the resistance at low $T$, including variable range hopping (VRH) [32], weak localization (WL) [35], the intrinsic properties of a Tomonaga-Luttinger Liquid (TLL) [4], or a thermally-activated gap [36] (see Appendix A for further discussion). Previous work on $Na_2Mo_6Se_6$ found that the insulating instability was well-described by a q1D VRH model [32], presumably due to Na vacancy-induced disorder. In contrast, both the VRH and WL models fail to reproduce our transport data in $Rb_2Mo_6Se_6$ [see Appendix A, Fig. 6 and Fig. 7], thus indicating that disorder does not play a dominant role in driving the insulating instability. This is consistent with the lower number of guest ion vacancies in as-grown $Rb_2Mo_6Se_6$, due to the larger mass of the Rb atom relative to Na and hence its reduced ionic mobility during the crystal growth process. Our resistivity upturn also cannot be fitted by a power-law model (see Appendix A, Fig. 8), implying that we are not observing the intrinsic behavior of a half-filled TLL with repulsive interactions [4]. The best interpretation of our data is found using a thermally-activated transport model, which corresponds to the opening of a gap $E_g$ in the density of states at the Fermi energy. However, an Arrhenius fit with constant $E_g$ is unable to reproduce our results. Instead, we model $R(T)$ with a $T$-dependent gap $E_g(T)$, as pioneered in Li$_{0.9}$Mo$_6$O$_{17}$ [36]. Using $R(T) = R_0 \exp[E_g(T)/2k_BT]$, where $R_0$ is defined as the resistance minimum $R_{\min}$, we may hence determine $E_g(T)$ as:

$$E_g(T) = 2k_BT (\ln R(T) - \ln R_{\min}) \quad (1)$$

The $T$-dependence of this gap magnitude is shown in Fig. 2(a) up to $P$ = 32 GPa. Above this pressure the analysis fails, since $T_{R\min}$ becomes indistinguishable from an impurity/defect scattering-induced saturation in $R(T)$. $T_{Egmax}$ can be determined up to 24.2 GPa, above which the analysis is numerically unstable.

Upon reducing $T$, the gap $E_g(T)$ starts to open at $T_{R\min}$ and reaches a maximum $E_{gmax}$ at $T_{Egmax}$. On further decreasing $T$ towards zero, $E_g(T)$ is observed to fall (though remains non-zero). For all $T < T_{R\min}$, application of pressure suppresses the gap magnitude $E_g(T)$. This is illustrated by Fig. 2(b), in which $E_{gmax}$ and $E_g$ ($T$=2 K) both decay approximately monotonically with increasing pressure. In contrast, $T_{Egmax}$ and $T_{R\min}$ first rise and then decay with increasing $P$ [Fig. 2(c)], both developing abroad maximum before falling above $P$ = 8.8 GPa.

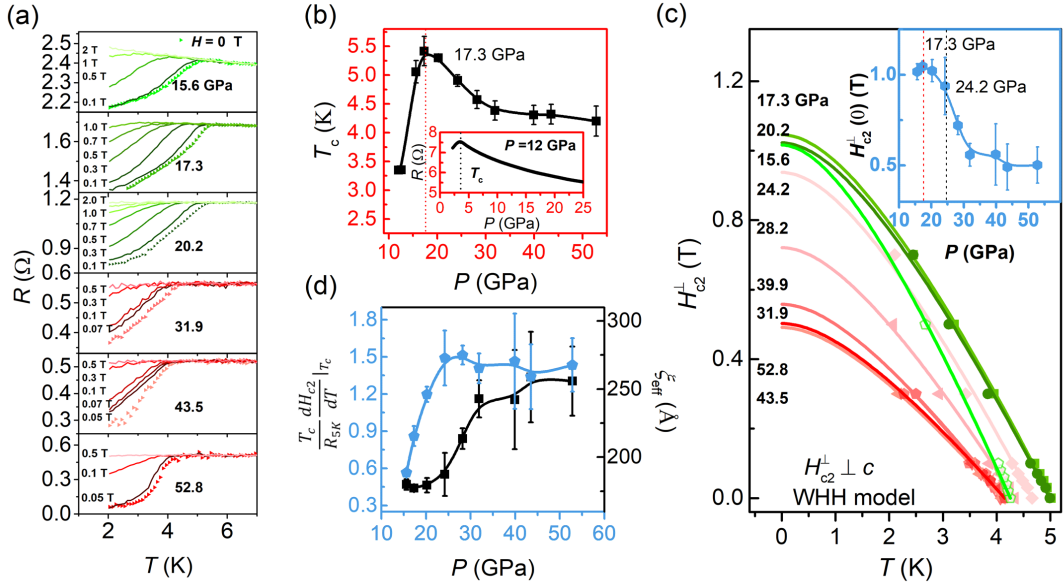

FIG. 3. (a) Electrical resistance of Rb$_2$Mo$_6$Se$_6$ versus temperature showing the SC transition for different values of the applied pressure: at zero field (black datapoints) and in various applied magnetic fields perpendicular to the $c$ axis (colored data). (b) Variation of the superconducting transition temperature $T_c$ with pressure. (c) Dependence of the upper critical field $H_{c2}^\perp$ with $T$ for $H \perp c$, at pressures varying from 15.6-52.8 GPa (datapoints). WHH fits of $H_{c2}^\perp(T)$ are also plotted (solid lines). The inset shows the pressure dependence of the extrapolated zero-$T$ upper critical magnetic field $H_{c2}^\perp(P)$. (d) Left axis: evolution of $\frac{T_c}{R_n}\frac{dH_{c2}}{dT}|_{T_c}$ with pressure, which is proportional

to the product of the dressed density of states in the superconducting phase $\Delta_0 \gamma$. Right axis: pressure dependence of the effective coherence length $\xi_{\text{eff}}$ (see text for detail).

We now address the SC phase, which can be identified as an abrupt drop in *R(T)* emerging at *P* = 12.0 GPa [inset of Fig. 3(b)]. The critical temperature $T_c$ is defined as the crossing of the linear regressions of *R(T)* above and below the transition (see Appendix B Fig. 9 for $T_c$ and error bar determination). Upon increasing the pressure $T_c$ initially rises steeply [Fig.1(c)], coinciding with the decline in $T_{\text{Egmax}}$ [Fig. 2(c)], before following a dome-shaped phase boundary with a maximum $T_c$ = 5.4 K at *P* =17.3 GPa. For all pressures, superconductivity can be quenched by applying a magnetic field $H \lesssim$ 2T [Fig. 3(a)]. Our pressure cell geometry obliges this field to lie perpendicular to the *c* axis of $Rb_2Mo_6Se_6$. The $H_{c2}^{\perp}(T)$ data which we extract from our *R(T)* curves can be extrapolated to *T* = 0 by fitting to the Werthamer-Helfand-Hohenberg (WHH) model[37] (Fig. 3c and Appendix B, Fig. 10):

$$\ln\frac{1}{t} = \sum_{n=-\infty}^{\infty} \frac{1}{|2n+1|} - [|2n+1| + \frac{\bar{h}}{t} + \frac{(\alpha \bar{h}/t)^2}{|2n+1|+(\bar{h}+\lambda_{SO})/t}]^{-1} \qquad (2)$$

where $t = T/T_c$, $\alpha = \frac{\mu_B H_{c2}(T=0, \alpha=0)}{\Delta_0}$ is the Maki parameter (describing paramagnetic limiting), $\bar{h} = -(4/\pi^2) H_{c2}^{\perp} / \frac{dH_{c2}^{\perp}}{dT}|_{t=1}$ is the reduced magnetic field, $\mu_B$ is the Bohr magneton, $\Delta_0$ is the zero-*T* superconducting gap, and $\lambda_{SO}$ is the spin-orbit coupling. Since we cannot cool below ~ 0.35 $T_c$, we simplify this process by fixing $\lambda_{SO}$ at zero and employing $\frac{dH_{c2}^{\perp}}{dT}|_{t=1}$ as a fitting parameter due to the sparse raw datasets. The resulting values of $H_{c2}^{\perp}$ (*T*=0, *P*) are shown in the inset of Fig. 3(c); their errors are determined by fits to $H_{c2}^{\perp}(T \pm \delta(H))$, where $\delta(H)$ are the errors in $T_c(H)$ deduced from our *R(T)* curves (Appendix B, Figs. 9, 10). Two distinct regions are immediately visible. Below 24.2 GPa, superconductivity coexists with the insulating instability: here $H_{c2}^{\perp}$ is large and weakly suppressed by pressure. However, at 24.2 GPa, a sudden drop in $H_{c2}^{\perp}$ occurs, followed by a strong suppression with rising pressure. Above 30 GPa, the pressure dependence of $T_c(P)$ ~ 4.1 K and $H_{c2}^{\perp}(T=0,P)$ ~ 0.5 T only evolve weakly. This contrasts starkly with the rise in, both $H_{c2}^{\perp}$ and $T_c$ at lower pressure, which is coincides with the onset of the insulating instability. $T_c(P)$ peaks at 17.3 GPa, but this dome-like behaviour is not reproduced in $H_{c2}^{\perp}(P)$. As we argue below, the lack of a clear dome in $H_{c2}(P)$ likely originates from a combination of increased anisotropy, emergent heterogeneity and enhanced coupling when superconductivity coexists with the insulating instability at low pressure.

Along with the transition temperature $T_c$, the shape of the superconducting transition also changes with pressure. Fig. 4 shows the resistivity curves in detail and analyses the width $\Delta T_c$ of the transition. At the lowest pressure where SC is observed, the onset of SC is followed by a gentle reduction of resistivity and hence a large value of $\Delta T_c$. With increasing pressure, the transition rapidly sharpens up, with a plateau between 25 and 42 GPa and a final further sharpening at the highest measured pressure of 52.8 GPa.

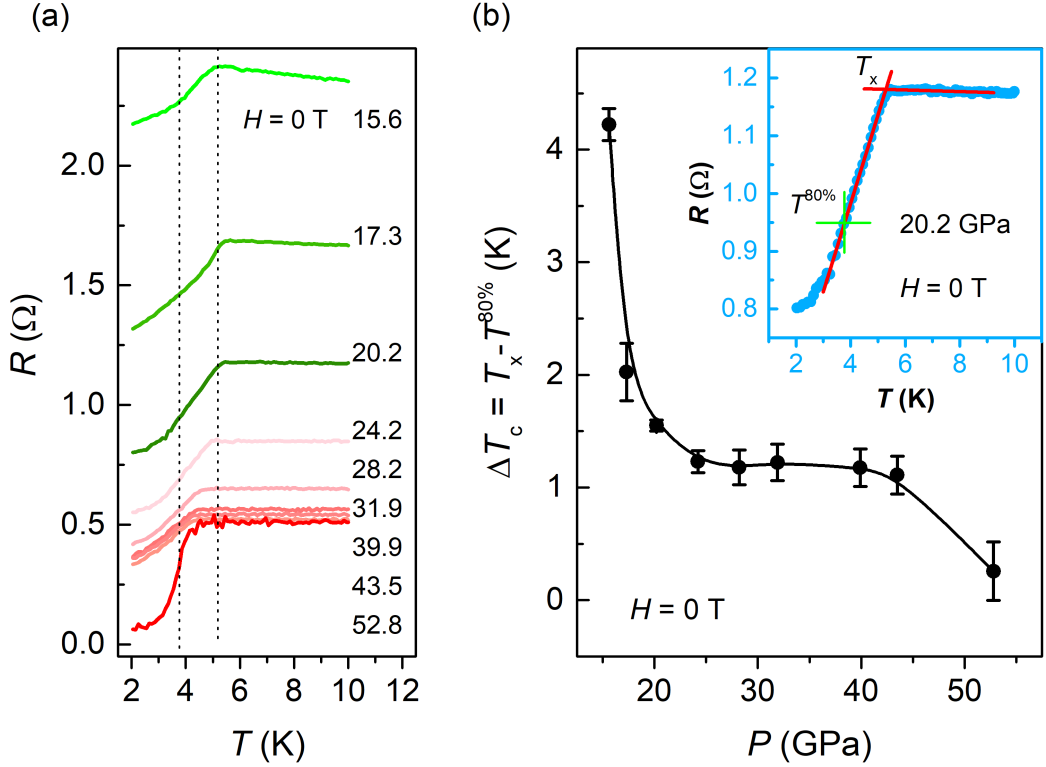

FIG. 4. (a) The SC transition of $Rb_2Mo_6Se$ at different pressure and zero field ($H = 0$ T). Dashed lines are guides to the eye. (b) Zero-field superconducting transition widths as a function of pressure. The insert exemplifies the method of determining $\Delta T_c = T_x - T^{80\%}$. We derive $T^{80\%} = (0.8bT_x + 0.8a - a')/b'$ from the same coefficients $a$, $b$, $a'$, and $b'$ as explained above and calculate $\Delta T_c = T_x - T^{80\%}$.

# Discussion

## Peierls-like Gapped Transport

$M_2Mo_6Se_6$ have been speculated to lie close to a Peierls-type instability due to their low $T$ metal to insulator transition, which should be tunable by pressure or uniaxial strain [13, 20, 38, 39]. However, neither a clear transition, nor a periodic lattice distortion have been observed (see e.g. our x-ray diffraction data at $T = 20$ K in Appendix C, Fig. 11, which show no deviation from regular lattice diffraction). In the present work, we demonstrate the emergence of a $T$-dependent gap $E_g(T)$, extracted from a thermally-activated model, which corresponds to a depletion of the density of states at the Fermi energy. On further decreasing $T$, the gap reaches a maximum $E_{gmax}$ and then falls again towards the lowest measured $T = 2$ K. We relate the initial rise in $T_{Rmin}$ between 0 and 8.8 GPa to the increased quenching of 1D fluctuations within a Lee-Rice-Anderson scenario framework [40]. This scenario is a model for the study of electron dynamics in the presence of long but finite ranged density wave order. Electrons are coupled to quasistatic (relevant frequencies less than $k_BT$) order parameter fluctuations, resulting in suppression of the single-particle density of states at low energies, a phenomenon sometimes referred to as a "pseudogap" [40, 41]. Within the pseudogap phase, the CDW competes with low dimensional quantum fluctuations: these act to suppress the Peierls transition temperature $T_p$ well below its mean-field value $T_{MF}$, with the CDW energy gap (the order parameter) fluctuating in time and space between these temperatures. Rb$_2$Mo$_6$Se$_6$ displays a nontrivial combination of low-dimensional fluctuation effects and a dynamical CDW with possible momentum dependence. However, at zero/low pressures, such fluctuations will suppress the onset temperature of the insulating instability. Initially, q1D fluctuations are suppressed with pressure, which extends $T_{Egmax}$ and $T_{Rmin}$ to higher $T$ with a maximum at 8.8 GPa. Above 8.8 GPa the insulating state starts to collapse, as the pressure-enhanced three-dimensionality (3D) of the system takes over, and $T_{Rmin}$ and $E_g$ both decrease [39]. This dimensional crossover is caused by pressure reducing the chain separation, thus favoring interchain coupling [25]. Above 31.9 GPa, the q1D to 3D transition is complete and the insulating behavior has fully disappeared.

## Coexistence of Superconductivity and Gapped Behaviour

In the pressure range between 15.6 and 24.2 GPa, a low-$T$ uprise of resistivity and a superconducting transition are observed together. To explore the coexistence mechanism in more detail, we employ the dirty limit expression for the Maki parameter: $\alpha = \frac{3e^2\hbar\gamma\rho_n}{2m\pi^2 k_B^2}$ [37, 42], where $\gamma$ is the dressed density of states at the Fermi level and $\rho_n$ is the normal-state resistivity. Setting this equivalent to $\frac{\mu_B H_{c2}(T=0,\alpha=0)}{\Delta_0}$, we recall that $H_{c2}(T=0,\alpha=0) \propto T_c \frac{dH_{c2}}{dT}|_{T_c}$. By evaluating $\frac{T_c}{R_n}\frac{dH_{c2}}{dT}|_{T_c}$ (where $R_n$ is the resistance measured immediately above $T_c$), we hence obtain a parameter proportional to $\Delta_0\gamma$ whose pressure evolution we plot in Fig. 3(d) (Left axis). The steep decay in $\Delta_0\gamma$ in the coexistence region indicates that the insulating instability is indeed removing states from the Fermi energy, but the fact that $T_c(P)$ continues to rise until 17.3 GPa suggests that $\Delta_0$ simultaneously increases to partially compensate the fall in $\gamma$. This behavior is reminiscent of the disorder-enhanced pairing observed in Na$_{2-\delta}$Mo$_6$Se$_6$ [32].

A common feature of superconductivity on the border of an insulating state is an emergent heterogeneity in the pairing interaction [43]. To explore the possibility of this phenomenon in Rb$_2$Mo$_6$Se$_6$, we track the pressure dependence of the SC transition width $\Delta T_c$, shown in Fig. 4. Although our crystals do not display true zero resistance, even at the highest applied pressure of 52.8 GPa, we can nevertheless devise a robust protocol to estimate $\Delta T_c(P)$ [Fig. 4(b)]. Low-dimensional phase fluctuations could plausibly influence the superconducting state at lower pressures (below 24 GPa), since the zero field and $P$-dependent SC transition appears to be widened here. It is difficult to distinguish between the influences of such fluctuations and any nascent inhomogeneity, since both will increase (broaden) $\Delta T_c$. Still, we rather think that inhomogeneity (due to microscopic spatial separation between superconducting and insulating zones) has a greater effect on superconductivity than fluctuations, because the $\Delta T_c$ curve only rises steeply below 20 GPa whereas $\xi_{\text{eff}}(P)$ (which determines the coherence volume and hence the phase stiffness [44]) already starts to drop below 30 GPa. A microscopically heterogeneous distribution of the insulating phase in $Rb_2Mo_6Se_6$ could be expected to induce electron localization and hence pairing enhancement due to the q1D geometry.

## Anisotropic Coherence of the Superconducting Condensate

By incorporating an anisotropic effective mass in Ginzburg-Landau theory [45], one can conveniently describe the anisotropy in the upper critical field in q1D superconductors. For field applied perpendicular to the high symmetry axis. $H_{c2}^{\perp} = \Phi_0/2\pi\xi_{//}\xi_{\perp}$ is controlled by the (long) coherence length parallel to the crystallographic chains, $\xi_{//}$, as well as the (short) transverse coherence length, $\xi_{\perp}$ [46]. Usually, one would first determine $\xi_{\perp}$ by applying a magnetic field parallel to the chains, then independently extract $\xi_{//}$ from $H_{c2}^{\perp}$; unfortunately, this is not possible in our experiments since our pressure cell is too large to be rotated within the magnet bore. We therefore define an effective coherence length $\xi_{\text{eff}} = \sqrt{\xi_{//}\xi_{\perp}}$ and plot its pressure dependence in Fig. 3(d) of right axis. Our anisotropic coherence lengths are linked to the electron effective masses via the relation $\xi_{//}/\xi_{\perp} \propto \sqrt{m_{\perp}^*/m_{//}^*}$. The S-shaped non-linear trend which we observed in $\xi_{\text{eff}}(P)$ is hence inconsistent with a simple linear evolution of the effective mass anisotropy with pressure. Instead, our results suggest that coexistence of superconductivity with the insulating phase imposes a limit on the coherence length. Moreover, since the (clean-limit) BCS coherence length is controlled by the ratio $\frac{v_F}{\Delta_0}$ and the Fermi velocity $v_F$ is approximately constant across the Fermi surface of an ideal q1D metal, a momentum-dependent gap which gradually depletes the Fermi surface cannot be solely responsible for the observed reduction in $\xi_{\text{eff}}(P)$. Our data indicate that increased scattering (due to spatial inhomogeneity) and/or an anomalous enhancement in the pairing interaction must coincide with the onset of insulating behavior.

## Phase Diagram

Three clear phases are thus distinguished in the phase diagram of $Rb_2Mo_6Se_6$ (Fig. 5): an anisotropic metal, an insulator and a superconductor. The most interesting zone of the phase diagram lies at 12.6 GPa < $P$ < 24.2 GPa, where the SC state forms a $T_c(P)$ dome in coexistence with the thermally-activated $E_g(T)$ gap. This calls for further experiments to verify the nature of the CDW (and its fluctuating order parameter) responsible for $E_g(T)$. Although a pseudogap should open in the density of states, no static periodic lattice modulation or spectroscopic gap have been directly observed to date. In future, spatial and momentum-resolved spectroscopy would be helpful to verify the amplitude and possible heterogeneity of $E_g$, hence revealing the precise interaction between superconducting and insulating states.

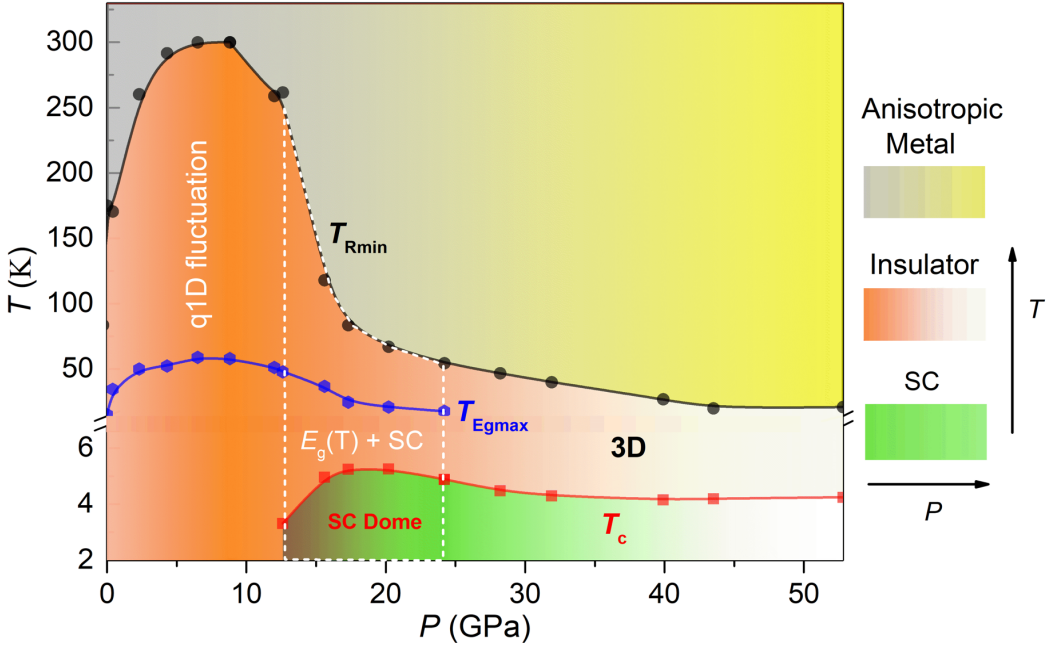

FIG. 5. Phase diagram in $Rb_2Mo_6Se_6$ as a function of pressure and temperature. The grey to yellow shading represents the evolution from q1D to 3D metal with increasing pressure. The green shading is proportional to $\Delta T_c$. At low $T$ and low $P$ (< 12 GPa), q1D fluctuations are dominant. In the 12.6-24.2 GPa pressure range, SC appears at low $T$ and traces out a dome in coexistence with the $T$-dependent gap $E_g(T)$. Upon further increasing $P$ (> 24.2 GPa), a rise in $\xi_{\text{eff}}(P)$ indicates a crossover to 3D behavior, with SC surviving up to the highest pressure applied.

## Conclusions

Pressure-dependent resistivity measurements have provided us with a rare opportunity to trace the ground state evolution of a quasi-one-dimensional crystal with continuously decreasing anisotropy. In the extreme 1D limit, we find an insulating state displaying transport characteristics which are consistent with a CDW whose onset temperature is suppressed by fluctuations. Raising the pressure increases the electronic dimensionality of the system, briefly stabilizing the insulating state, before $T_{\text{Rmin}}$ starts to collapse above 8.8 GPa. A superconducting ground state then emerges at 12 GPa and survives up to our highest accessible pressure (52.8 GPa). The superconducting region of the phase diagram can be divided into two zones: a conventional, weakly-anisotropic state at high pressures ($P \gtrsim 30$ GPa), and an unconventional regime at lower pressures in which both $T_c$ and $H_{c2}^{\perp}$ increase due to coexistence with the insulating phase.

Our results highlight two outstanding questions which merit further attention from techniques beyond bulk DC transport. Firstly, the anomalous suppression of the insulating gap at low temperature (which we infer from our *R(T)* curves) defies any simple explanation within a density wave scenario. Secondly, the enhancement of $T_c$ and $H_{c2}^{\perp}$ occurs despite a concomitant depletion in the density of states, thus indicating a strengthening of the pairing interaction by an as-yet unclear mechanism. We speculate that both these effects may be consequences of spontaneous microscopic phase separation in nominally homogeneous q1D crystals: an unusual phenomenon which may be relevant for other low-dimensional systems with competing instabilities. Future experiments should address this possibility, ideally using spatially-resolved spectroscopic and/or structural probes.

# ACKNOWLEDGMENTS

This work was supported by the Helmholtz-Office of China Postdoc Council (OCPC) Postdoc Program and National Nature Science Foundation of China (Grants No. U1930401, No. 51527801, and No. 51772184) and National Key R&D Program of China (Grant No. 2017YFA0403401). We wish to thank Dmitry Chernyshov for data taking and deep discussions and the Swiss-Norwegian Beamlines (SNBL) at the European Synchrotron Radiation Facility (ESRF) for access, which contributed to the results shown in the Appendix.

# Appendix

## A) Possible Models for the divergent resistivity at low-temperature

Here we discuss four candidate models for the *T*-dependent resistivity which could conceivably describe our data, but eventually fail to give a reasonable fit. We therefore consider the underlying physical mechanisms to be irrelevant or sub-dominant in $Rb_2Mo_6Se_6$.

**Variable Range Hopping (VRH):**
The VRH model applies to a strongly disordered electron system, in which charge transport occurs by hopping between nearby localized states:
$$\rho(T) = \rho_0 \exp[(T_0/T)^{1/(1+d)}] \quad (A1)$$
where $T_0$ is the effective localization temperature, which describes both the hopping length and activation energy, and $d$ is the dimensionality of the system. The fitted data clearly deviate from the observed resistance, as illustrated in both the linear and logarithmic scale plots shown in Fig. 6.

**Weak Localization:**
Weak localization is generally encountered in less-disordered low-dimensional systems. It is a disorder-induced quantum interference phenomenon resulting from enhanced backscattering of delocalized electrons, creating a logarithmically-divergent resistivity:
$$\rho(T) = \rho_0 + \rho_1/\ln(T) \quad (A2)$$
where $\rho_0, \rho_1$ are constants. Our attempt to fit the low-temperature transport data using this model also does not result in convergence (Fig. 7).

**Tomonaga-Luttinger Liquid (TLL):**
We consider a TLL with strongly repulsive electron-electron (e⁻-e⁻) interactions, i.e. a Luttinger parameter $K_\rho \ll 1$. In this case, a power-law suppression in the density of states $N(E) \propto E^\alpha$ creates a pseudogap at the Fermi level and:
$$\rho(T) \propto T^\alpha \quad (A3)$$
with $\alpha = 4n^2 K_\rho - 3 < 0$, and commensurability $n = 1$ for half-filling. Our data are manifestly not compatible with this scenario (Fig. 8).

**Activated behavior:**
Any phase transition which opens a hard gap in the density of states at the Fermi energy will exhibit thermally-activated transport following the Arrhenius equation:
$$\rho(T) \propto \exp[E_g/2k_BT] \quad (A4)$$
A conventional density wave (DW) which gaps the entire Fermi surface will fall into this category. The corresponding phase transition would also create a discrete jump in $\rho(T)$ rather than a continuous exponential divergence. Using Arrhenius plots of log($R$) versus $1/T$, we have verified that no stable gap value over any reasonable range of temperature matches the data.

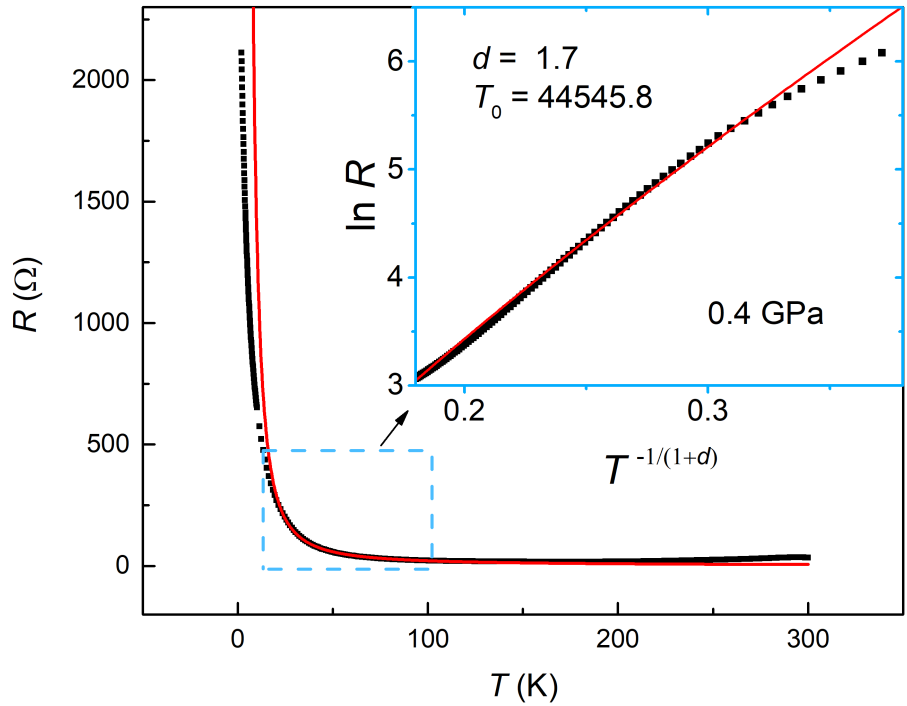

FIG. 6. Variable range hopping (VRH) fits of our low temperature $R(T)$ data at 0.4 GPa. The fitting results at higher $T$ visibly deviate from the real resistance values. The inset (on logarithmic axes) shows the zooming in the rectangular box on linear axes, and highlights the case of $\nu = 1/(1+d)$ with $d = 1.7$, which clearly demonstrates that VRH is not responsible for our low temperature resistivity upturn.

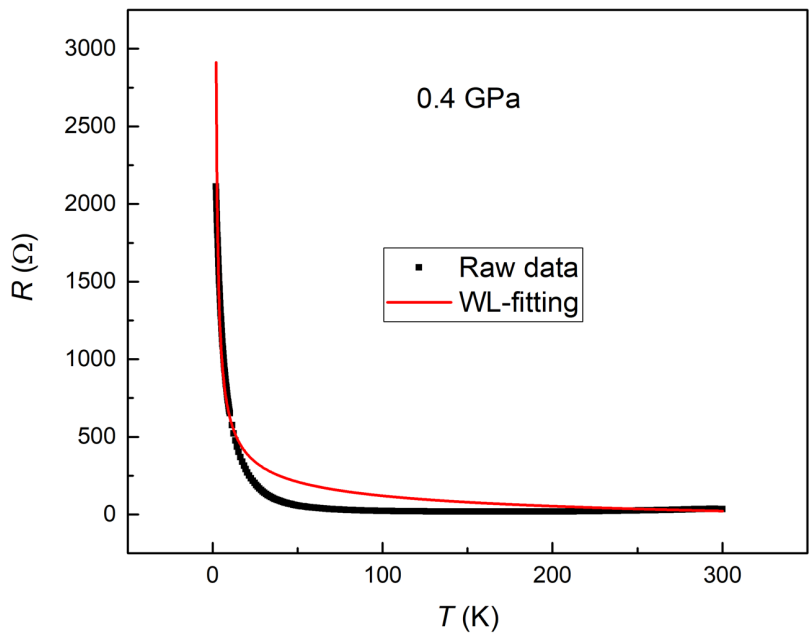

FIG. 7. The weak localization fitting results at 0.4 GPa fail to converge and the best fit shows obvious deviations from the data.

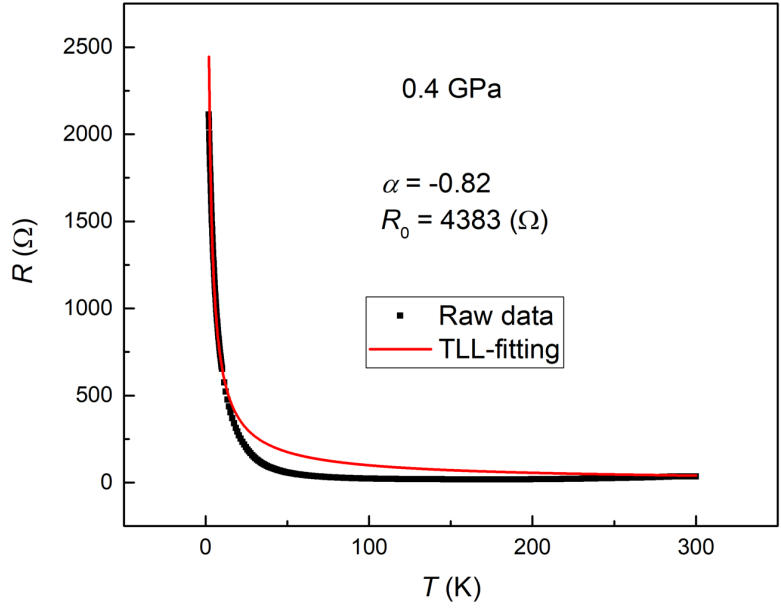

FIG. 8. The Tomonaga-Luttinger Liquid (TLL) power-law model cannot reproduce the data.

**B) Method of $T_c$ determination**

To arrive at an unambiguous method of determining the SC transition temperature $T_c$, we identify ranges of linear $R(T)$ above and below the kink at $T_{onset}$ which reveals the onset of SC. We extract parameters $a$ and $b$ by linear regression $R(T)=a+bT$ above $T_{onset}$, and $a'$ and $b'$ by $R(T)=a'+b'T$ below $T_{onset}$. The SC temperature $T_c$ is then determined as the intersection of the two linear fits. This $T_c \equiv T_x$ is used and discussed for all data at zero magnetic field. An identical method is employed for all $R(T)$ datasets in our analysis of the quenching of SC at non-zero magnetic field. Here, we determine $T^{96\%}(H)$ =$[(0.96*a+0.96*b*T)-a']/b'$. These $T^{96\%}(H)$ are used to estimate zero-$T$ upper critical fields via our WHH fits. Their uncertainty $\Delta T^{96\%}(H)$ is given by:

$$\delta(H) = T^{96\%}(H)\sqrt{(\Delta a'/(a-a'))^2 + (\Delta b'/(b-b'))^2} \qquad (B1)$$

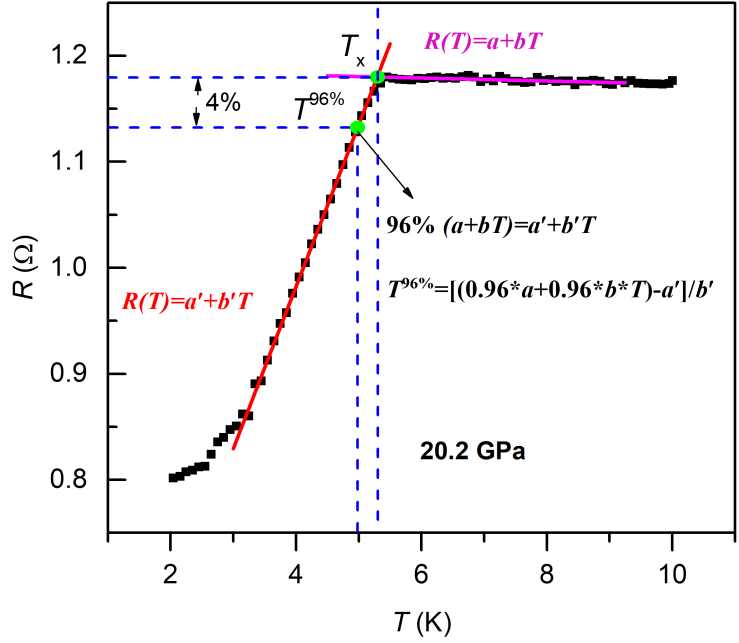

FIG. 9. Example of our $T_c(H)$ determination method and its error bar calculation at 20.2 GPa with $H = 0$.

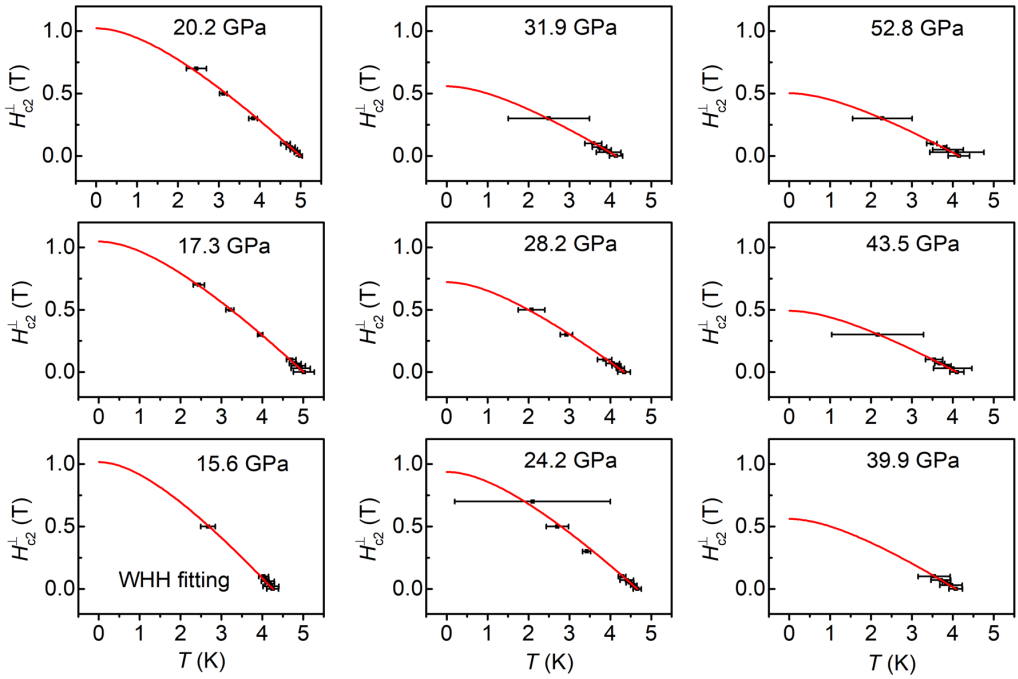

FIG. 10. Temperature dependence of the upper critical field $H_{c2}^{\perp}$ [black datapoints, extracted using $T_c^{96\%}(H)$] and corresponding WHH fits to $H_{c2}^{\perp}(T)$ (red lines) for various pressures. Error bars $\delta(H)$ are included for each pressure.

## C) Non-observations of a periodic lattice distortion in x-ray diffraction

Single crystal x-ray diffraction data at ambient pressure have been obtained at the Swiss-Norwegian Beamline at the European Synchrotron Radiation Facility (ESRF). The data were analyzed and re-assembled into plots of intensity in arbitrary planes of reciprocal space. An example of such reconstructed data is shown in Fig. 11. The full data set extends down to $T = 18$ K and intense scrutiny in various cuts of the data set failed to show any diffraction features besides the regular crystal lattice diffraction.

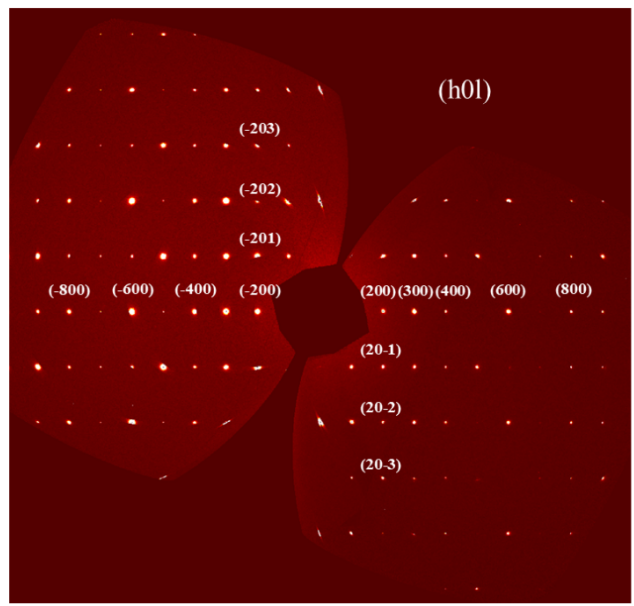

FIG. 11. Reconstructed single crystal diffraction (h 0 l) patterns for $Rb_2Mo_6Se_6$ at ambient pressures and low temperature ($T = 20$ K).